# Quantum Mechanical Process of Carbonate Complex Formation and Large Scale Anisotropy in the Adsorption Energy of $CO_2$ on Anatase $TiO_2$ (001) Surface

Shashi B. Mishra,[1] Aditya Choudhary,[1] Somnath C. Roy,[2] and B. R. K. Nanda[1,*]

[1]*Condensed Matter Theory and Computational Lab, Dept. of Physics, IIT Madras, Chennai, India*

[2]*Environmental Nanotechnology Lab, Dept. of Physics, IIT Madras, Chennai, India*

## ABSTRACT

Adsorption of $CO_2$ on a semiconductor surface is a prerequisite for its photocatalytic reduction. Owing to superior photocorrosion resistance, nontoxicity and suitable band edge positions, $TiO_2$ is considered to be the most efficient photocatalyst for facilitating redox reactions. However, due to the absence of adequate understanding of the mechanism of adsorption, the $CO_2$ conversion efficiency on $TiO_2$ surfaces has not been maximized. While anatase $TiO_2$ (101) is the most stable facet, the (001) surface is more reactive and it has been experimentally shown that the stability can be reversed and a larger percentage (up to ~ 89%) of the (001) facet can be synthesized in the presence fluorine ions. Therefore, through density functional calculations we have investigated the $CO_2$ adsorption on $TiO_2$ (001) surface. We have developed a three-state quantum-mechanical model that explains the mechanism of chemisorption, leading to the formation of a tridentate carbonate complex. The electronic structure analysis reveals that the $CO_2$-$TiO_2$ interaction at the surface is uniaxial and long ranged, which gives rise to anisotropy in binding energy (BE). It negates the widely perceived one-to-one correspondence between coverage and BE and infers that the spatial distribution of $CO_2$ primarily determines the BE. A conceptual experiment is devised where the $CO_2$ concentration and flow direction can be controlled to tune the BE within a large window of ~1.5 eV. The experiment also reveals that a maximum of 50% coverage can be achieved for chemisorption. In the presence of water, the activated carbonate complex forms a bicarbonate complex by overcoming a potential barrier of ~0.9 eV.

* Electronic address: nandab@iitm.ac.in

# I. INTRODUCTION

Recycling of carbon dioxide by converting it into hydrocarbons has been suggested as an energetically efficient method for the mitigation of $CO_2$ concentration [1–9]. The direct $CO_2$ reduction requires dissociation of the C = O bond that costs an energy of 750 kJ/mol [10], which is thermodynamically difficult to achieve. Therefore, a catalytic reduction, which involves several low-energy multistep processes *via* protonation and electron capture, is more favorable. To develop an efficient catalyst, long-term stability and enhanced $CO_2$ adsorption capability are desired. In the case of photocatalysis, anatase $TiO_2$ is one of the preferred choices because of its superior chemical stability and appropriate band-edge position (with respect to redox potential) [11–19]. While a lot of work has been reported to improve the photocatalytic ability [20–22], the $CO_2$ – $TiO_2$ interaction mechanism has not been explored.

From a thermodynamical analysis of an experimental study, Koppenol and Rush [23] reported that the catalytic reduction of $CO_2$ initiates with an electron capture that leads to the formation of $CO_2^-$ ions. But the inert nature of $CO_2$ along with its positive electron affinity (~ 0.6 eV) [24] results in a reduction potential of -1.9 V (vs standard hydrogen electrode) for a $CO_2 + e^- \rightarrow CO_2^-$ reaction [25]. Such a high potential hinders the transfer of electrons, thus making it a rate determining step [26,27]. The formation of $CO_2^-$ anions on a pure $TiO_2$ surface is also evident by the vibrational spectroscopic technique which confirms the charge transfer between $CO_2$–$TiO_2$ [28].

Several experimental [28–30] and theoretical studies [27,31–36] have been carried out to study the $CO_2$ adsorption on the most stable (101) surface of anatase $TiO_2$. However, in two seminal works [37,38], it has been experimentally shown that the stability can be reversed and a larger percentage (up to 89%) (001) facet can be synthesized in the presence fluorine ions. Further, Han *et al.* [38] and several others [21,39–42] have experimentally shown that the (001) facet substantially enhances the photocatalytic efficiency.

In the context of a stable (101) surface, several first-principles calculations have been performed using both cluster and slab models to study the activation of the $CO_2$ on anatase $TiO_2$ surfaces [31–35,43,44]. Relying upon the energetics and charge analysis, He *et al.* have reported that the conversion of $CO_2$ to formic acid on $TiO_2$ (101) surface begins with the initial activation of $CO_2$ *via* one electron transfer to form the $CO_2^-$ anion with an activation barrier of 0.87 eV [33,34], and also observed that this step plays a crucial role in determining the efficiency of conversion. Hence, to improve the catalytic performance on $TiO_2$, the adsorption and activation of $CO_2$ plays a crucial role.

Considering the adsorption on (101) surface, Sorescu *et al.* [44] and others [32–35] have estimated the binding energy (BE) close to -0.5 eV. It suggests that $CO_2$ in this case is physisorbed where both surface and adsorbate are negligibly deformed. On the other hand, the density functional theory (DFT) calculations for adsorption on (001) surface, using clusters [32,45] and slabs [35,46], yield stronger BE (-1.1 to -1.4 eV). In this case, the formation of a carbonate complex and the deformed surface suggests chemisorption of $CO_2$ molecule. However, the quantum mechanical process leading to such complex formation has not been understood. Gaining insight into it is important to unravel the $TiO_2$ – $CO_2$ interaction and hence the charge transfer mechanism. We will see that the latter introduce binding energy anisotropies which are crucial for quantification of adsorption, activation and reduction of $CO_2$ molecules on the $TiO_2$ (001) surface.

For the development of economically viable technology, high $CO_2$ conversion efficiency is desired. In this context, a few experimental and theoretical studies have been carried out to investigate the effect of $CO_2$ concentration on the adsorption and reduction ability of the catalyst surface [35,44,47–49]. Through DFT calculations, Sorescu *et al*. have examined the role of $CO_2$ coverage on anatase (101) surface [44] and found that there is a negligible increase in the adsorption energy (~1 meV) when the coverage is increased from 10 to 25%. As mentioned earlier, $CO_2$ molecules are weakly adsorbed on the (101) surface and hence it is expected that the BE will not be as sensitive to the coverage. However, the (001) surface being highly reactive and since $CO_2$ molecules are strongly adsorbed on this surface, a deterministic investigation is required to develop a relation between coverage and adsorption energy.

In the present work, through DFT calculations (see Computational Details), we have developed a three state model to appropriately explain the $CO_2$–$TiO_2$ chemical interaction occurring at the twofold coordinated oxygen atom [$O_{2f}$ : Fig. 1(a)], which is the energetically preferred site for adsorption on the anatase $TiO_2$ (001) surface [Fig. 1(f)] [46]. A detailed analysis has been carried out on the basis of Löwdin charges and molecular-orbital theory to reveal the mechanism leading to the formation of tridentate carbonate complex on the host ($TiO_2$) surface. We have found that the inhomogeneously charged $TiO_2$ surface exert a restoring torque on the randomly oriented $CO_2$ molecules to make them aligned along the Ti–$O_{2f}$–Ti chain and eventually forms an identical carbonate complexes at the $O_{2f}$ site. From the coverage dependent adsorption analysis, we have observed that a maximum of 50% coverage can be achieved for chemisorption. Further, we have discovered the anisotropic behavior in BE, i.e. for the same coverage, the BE varies as a function of spatial distribution of the adsorbate. Such behavior, which has not been reported earlier, stems from the different chemical nature of Ti–$O_{2f}$–Ti and Ti–$O_{3f}$–Ti chains (three-fold coordinated O, $O_{3f}$ [Fig. 1(a)] along the [100] and [010] directions, respectively. Unlike the case of (101) surfaces, we observe that the variation in coverage can change the BE as much as 1.3 eV. This work, therefore, establishes the possibility to tune the BE and helps to achieve a better conversion efficiency by tailoring the distribution pattern of the $CO_2$ molecule on the $TiO_2$ surface. To mimic the real-world experiments, we have examined the coadsorption of $H_2O$ and $CO_2$. We find that in two independent processes, $H_2O$ dissociates to form two hydroxyl ions and $CO_2$ forms the activated carbonate complex. By overcoming a potential barrier of ~0.9 eV the latter breaks one of the neighboring hydroxyl ions and forms a bond with the freed H atom to develop a bicarbonate ($HCO_3$) complex.

## II. COMPUTATIONAL DETAILS

The DFT calculations are performed using ultrasoft pseudopotential and the plane wave basis sets as implemented in the Quantum ESPRESSO package [50]. Exchange-correlation potential is approximated through the Perdew-Burke-Ernzerhof general gradient approximation (PBE-GGA) functional [51]. Additionally, we have included the semi-empirical Grimme-D2 van der Waals correction [52]. The kinetic energy cutoff to fix the number of plane waves is taken as 30 Ry. The Brillouin zone integration is carried out using the tetrahedron method through a Monkhorst pack *k*-point grid [53]. A $4 \times 4 \times 1$ $k$ mesh is considered for the structural relaxation, while for electronic structure calculations, a denser $k$ mesh of $8 \times 8 \times 1$ is used. The convergence criterion for self-consistent energy is taken to be $10^{-6}$ Ry. All the structures are relaxed until the force on each atom is lower than 0.025 eV/Å. The charges on individual atoms are measured using Löwdin charge analysis. The structural and charge-density plots are generated using the visualization tool VESTA [54]. For some selective cases, the transition-state-theory based CI-NEB simulations [55] are carried out to examine the nature of transition between the initial and final configuration.

The ground state body centered tetragonal structure (space group $I4_1$; No. 141) with GGA optimized lattice parameters (a = 3.794 Å, c = 9.754 Å) is considered for the electronic structure calculations. Our optimized lattice parameters agree well with the earlier reported theoretical [56–58] and experimental [59] values. The surface is constructed using a slab model. A vacuum of 12 Å is found to be sufficient to avoid the interaction among the surfaces of the neighboring slabs. The surface energy is calculated using the following expression,

$$E_S = \frac{1}{2A}(E_{slab} - E_{bulk}), \tag{1}$$

where, $E_s$, $E_{slab}$ and $E_{bulk}$ represent the surface energy, total energy of the slab and $E_{bulk}$ represents the total energy per formula unit of bulk $TiO_2$ respectively, while $N$ stands for the total number of formula units present in a slab and $A$ is the surface area of the slab. We carried out surface energy convergence test and found that a six-layer-thick slab is sufficient to give well-converged adsorption energies. The BE of $CO_2$ molecule is calculated using the following equation:

$$BE = E_{TiO_2(001)/CO_2} - E_{TiO_2(001)} - E_{CO_2}, \tag{2}$$

where $E$ is total energy of the respective systems. The total energy of a $CO_2$ molecule is calculated by placing it in a large simple cubic unit cell ($a$ ~ 12 Å).

## III. RESULTS AND DISCUSSION

The atomic distribution of anatase $TiO_2$ (001) surface is made up of five-fold coordinated Ti ($Ti_{5f}$), two-fold coordinated O ($O_{2f}$) and three-fold coordinated O ($O_{3f}$) as shown in Fig. 1a. These three sites and the hollow position act as adsorption sites for the $CO_2$ molecule. As observed from Figs. 1(b)-1(e), upon adsorption there is a change in the geometry of the $TiO_2$ surface and the molecular structure of $CO_2$, which is significant only in the case of ontop $O_{2f}$. In this case, the O–C–O bond angle changes by 50.3° and the out-of-plane displacement of the $O_{2f}$ at the adsorption site is nearly 0.47 Å. For adsorption at other sites, the change in the bond angle is less than 5° and the corresponding out-of-plane displacement is less than 0.06 Å. To quantify the strength of adsorption, BE calculations are performed and the results are shown in Fig. 1(f). We find a BE of -1.66 eV for adsorption at $O_{2f}$, which implies a chemisorption; while for other sites, physisorption is observed with a weak BE of about -0.30 eV.

During the chemisorption of $CO_2$, the C atom develops a bond with $O_{2f}$ (bond length 1.32 Å). Concurrently, the C – O bond in $CO_2$ dilutes as their bond length increases from 1.18 to 1.28 Å. Finally, a carbonate complex is formed [32,35,46] which resembles an ideal carbonate anion where the O – C – O bond angle is 120° and the C – O bond length is 1.28 Å. Understanding the bonding mechanism will enable us to develop a phenomenology on adsorption as a function of both coverage and spatial distribution of the $CO_2$ molecule in a microscopic scale.

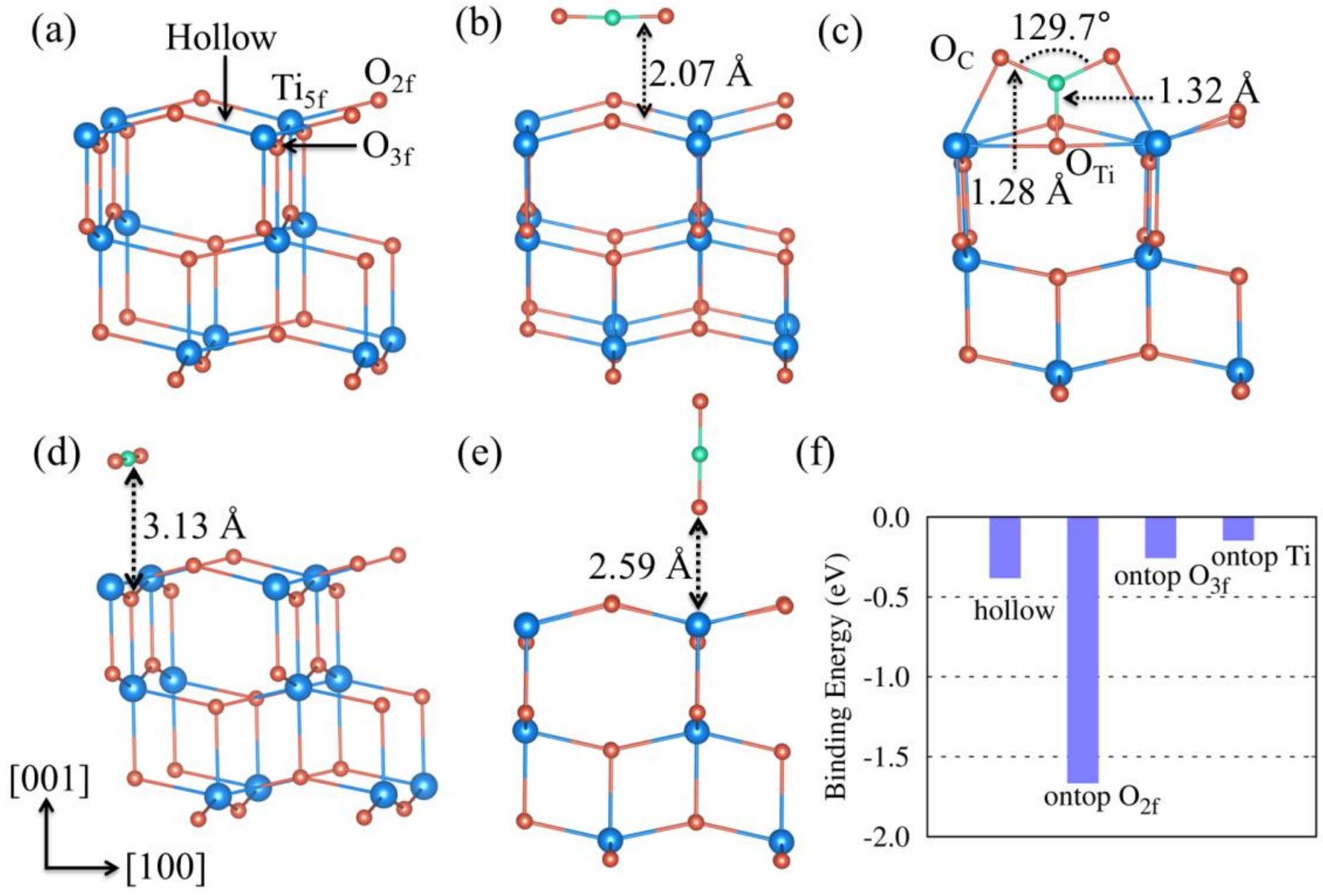

**FIG. 1.** (a) A 2 × 2 bare $TiO_2$ (001) surface showing the four possible adsorption sites. (b – e) The optimized structures after the adsorption of $CO_2$ over hollow, ontop $O_{2f}$, ontop $O_{3f}$, and ontop Ti positions respectively. (f) The binding energy for the corresponding sites.

## A. The mechanism of $CO_2$ adsorption and formation of carbonate complex

As a first step to understand the chemical bonding between the $CO_2$ molecule and the $TiO_2$ surface, we have performed the Löwdin charge analysis. The electronic charge on each site is obtained by summing the contribution of valence orbitals of the corresponding atom on each occupied band. Therefore, the net charge on the given site is the difference between electronic charge and total number of valence electrons of the corresponding atom. The results have been analyzed by comparing the charges on Ti and $O_{2f}$ of the $TiO_2$ surface and on C and O of the pure $CO_2$ molecule before and after adsorption. For convenience, henceforth we will refer to the adsorbing $O_{2f}$ site as $O_{Ti}$ and O atoms of $CO_2$ as $O_C$.

To gain in-depth insight into the nature and consequence of bonding and charge transfer, a three-state model is proposed as shown in Fig. 2. State I represents the configuration before adsorption i.e., the pristine $TiO_2$ (001) and linear $CO_2$ molecule are isolated, whereas state III represents the configuration after adsorption, which has a bent $CO_2$ molecule interacting with $O_{2f}$ of the deformed $TiO_2$ surface. State II represents a hypothetical configuration in which the $TiO_2$ surface and the $CO_2$ molecule are deformed as in state III, but without having any interaction between them. State II helps us to quantitatively estimate the instability of the deformed $TiO_2$ surface and bent $CO_2$ molecule as well as the charge redistribution accompanied by the deformation and bending. The eigenstates of each of the three states are further examined through the partial density of states (PDOS) shown in Fig. 4.

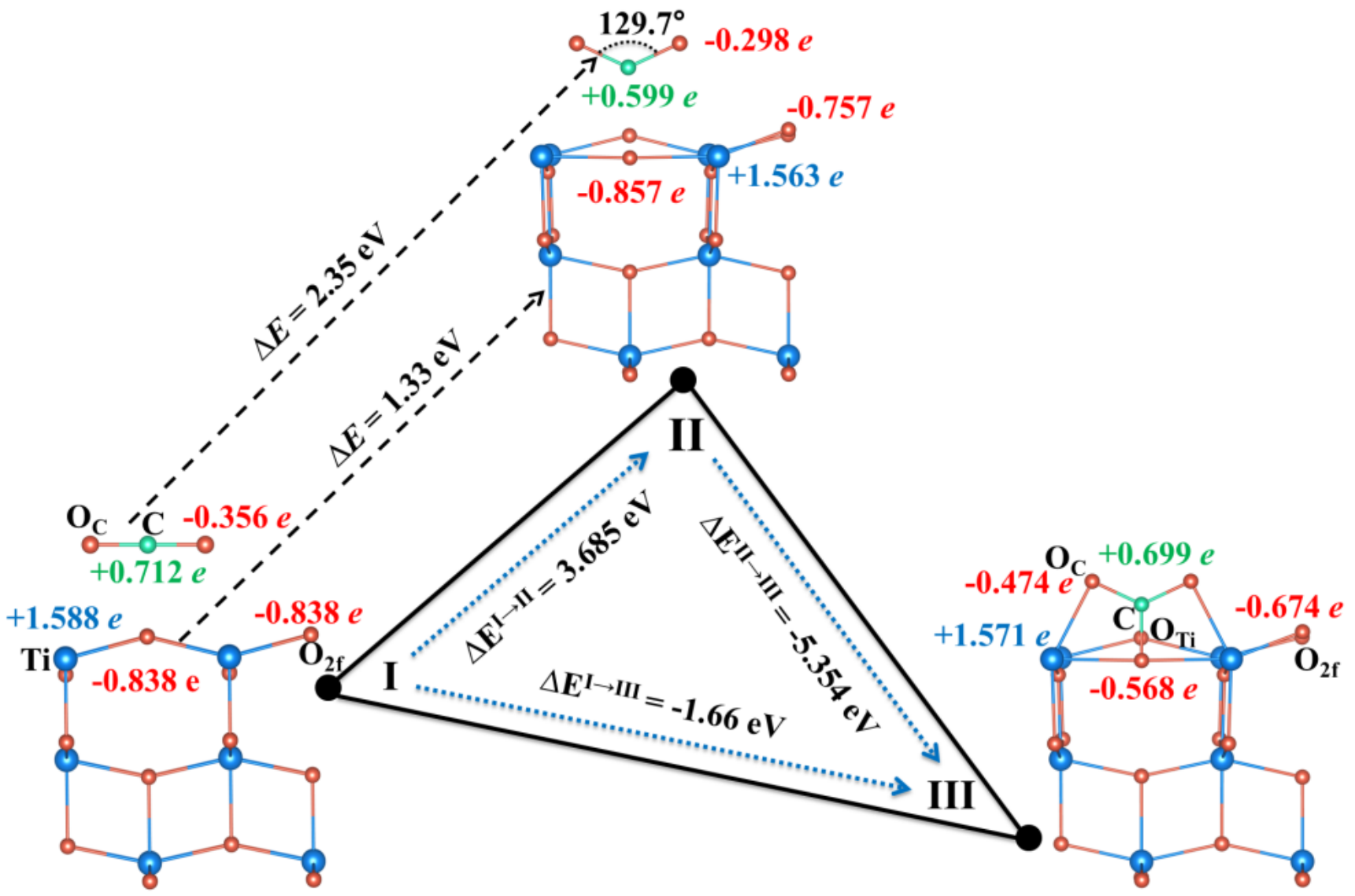

**FIG. 2.** The three-state model. State I and III represent the situation before and after adsorption at the $O_{2f}$ site respectively. The relative energies as well as site-specific Löwdin charges for each state are mentioned. State II is a geometrical replica of state III, but without any electronic interaction between the adsorbate and adsorbent. For convenience, the O atoms of $CO_2$ are referred as $O_C$ and the ontop $O_{2f}$ is referred as $O_{Ti}$. The results presented here are for the adsorption of a single $CO_2$ molecule on $2 \times 2$ $TiO_2$ surface. The charges on $O_{3f}$ (not shown in the figure) remain unchanged in all three states inferring that it is insensitive to adsorption.

As shown in Fig. 2, the charges of the Ti and $O_{2f}$ atoms on the bare $TiO_2$ surface (state I) are +1.588 *e* and -0.838 *e*, respectively. This resulted in a negatively charged surface with a net charge density of -$2.1\times10^{-3}$ $e/Å^2$, making the surface reactive towards the adsorbing molecules. On the other hand, an isolated $CO_2$ molecule has charge of +0.712 *e* and -0.356 *e* on the C and $O_C$ atoms respectively. As a result, a net attractive electrostatic force acts on the $CO_2$ molecule when it lies below a critical distance (estimated to be ~ 5.5 Å) from the $TiO_2$ surface. As the molecule approaches the surface, it experiences a site specific charge distribution and accordingly prefers to adsorb over a more negative $O_{2f}$ site to form a carbonate complex ($CO_3^{\delta-}$). A detailed description of the adsorption pathway for $CO_2$ is presented in Sec. IIIC.

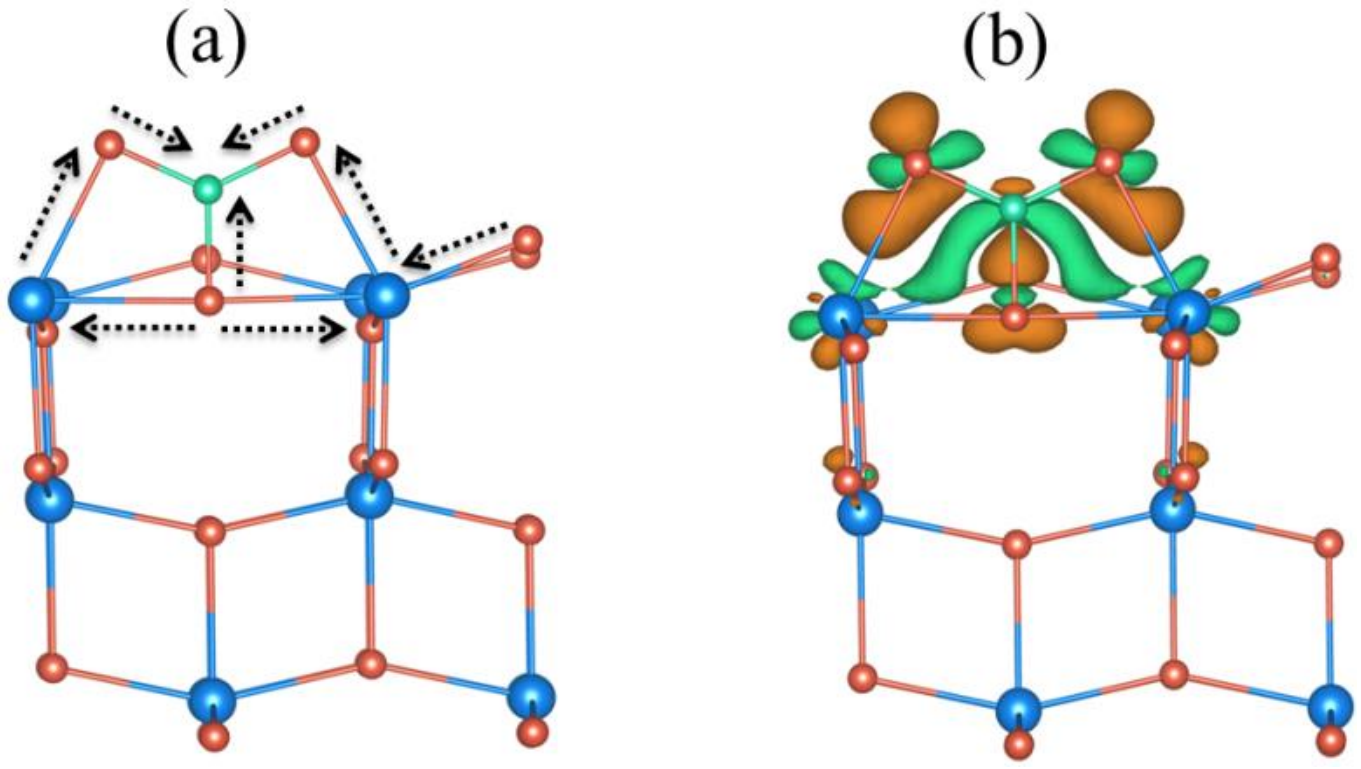

**FIG. 3.** (a) Charge transfer mechanism: The arrows indicate the direction of transfer of electrons during the formation of carbonate complex. (b) The charge density difference plot as calculated from the relation $\Delta\rho(r) = \rho_{TiO_2/CO_2}(r) - \rho_{TiO_2}(r) - \rho_{CO_2}(r)$. Here, $\rho_{TiO_2/CO_2}$ is the charge density of the state III, $\rho_{TiO_2}$ and $\rho_{CO_2}$ denotes the charge density of the $TiO_2$ and $CO_2$ molecule, respectively as present in state II. The isovalue is set to be 0.01 $e/Å^3$. Here, the brown and green color lobes represent the charge accumulation and depletion regions, respectively.

To gain an insight into the formation of carbonate complex ($CO_3^{\delta-}$), we split the adsorption process into two steps: I→II followed by II→III as shown in Fig. 2. Path I→II provides information about the redistribution of charges caused by the distortion alone, while path II→III helps to determine the charge transfer due to the bonding. Although distortion and bonding are simultaneous and interrelated phenomena during adsorption, treating them separately not only helps to analyze the individual effects but also assist in explaining the electron transfer between adsorbate and adsorbent. In state II, the charges on the C and $O_C$ atoms of the bent $CO_2$ molecule are found to be +0.596 *e* and -0.298 *e* respectively. When compared with the linear molecule, the sharing of electrons between C and $O_C$ atoms in the bent configuration is reduced by 0.116 *e*. This is attributed to the weakening of the covalent double bond due to deviation from axial geometry. As a consequence, the bent molecule becomes unstable by +2.35 eV. As far as the deformation of the $TiO_2$ surface is concerned, the downward displacement of $O_{Ti}$ increases the covalent interaction with the neighboring Ti due to the formation of linear Ti – $O_{Ti}$ – Ti geometry. This leads to a net gain of 0.02 electrons at the $O_{Ti}$ site. On the other hand, the deformation significantly weakens the axial interaction between Ti and $O_{2f}$, because of increase in the bond length, leading to a loss of 0.08 electrons from $O_{2f}$ to Ti. Although such deformation lowers electronic interaction energy, stronger ionic repulsion makes the system unstable.

The transition from state II to III initiates interactions between the deformed surface and the bent $CO_2$ molecule. As a result, new bonds, $O_{Ti}$ – C and $O_C$ –Ti, are developed and simultaneously some of the existing bonds, $O_{Ti}$ – Ti, $O_{2f}$ – Ti and $O_C$ – C, are weakened. This leads to a net loss of electrons by $O_{2f}$ and $O_{Ti}$ whereas there is a net gain of electrons by $O_C$. The direction of electron transfer during the adsorption process is schematically presented in Fig. 3(a). The net charge-density difference in the neighborhood of the adsorption site is shown in Fig. 3(b). The green and brown electron clouds indicate electron loss and gain, respectively as we move from state II to III.

It is important to compare the formed $CO_3^{\delta-}$ complex with the ideal $CO_3^{2-}$ ion. The former has a net charge of -0.817 *e*, while the latter has a net charge of -2 *e*. For an ideal carbonate ion, the charges on C and O, obtained from DFT calculations, are +0.589 *e* and -0.863 *e*, respectively; whereas the charges on C, $O_{Ti}$ and $O_C$ of formed complex are +0.699 *e*, -0.568 *e* and -0.474 *e*, respectively. Moreover, the newly formed C – $O_{Ti}$ and C – $O_C$ bond lengths are 1.32 Å and 1.28 Å, respectively, which are nearly the same as that of the C – O bond (~ 1.30 Å) of the ideal carbonate complex. The comparison of charges and bond lengths confirm the formation of carbonate-like complex.

The charge transfer process, shown in Fig. 3(a), suggests that the interaction between the adsorbate and adsorbent is not confined to the adsorbed site. The flow of charge from $O_{2f}$ towards the $CO_3^{\delta-}$ complex implies long range interaction of impurity ($CO_2$) states with the host ($TiO_2$) Bloch states. To understand the impurity-host interactions, we have analyzed the densities of states (DOS) plotted in Fig. 4 and mapped them to the chemical bonding of $CO_2$, $TiO_2$, and the composite. The chemical bonding of $CO_2$ is expressed through a general 12 × 12 Hamiltonian (*H*). With the basis set in the order $\{|C-s\rangle, |O1-s\rangle, |O2-s\rangle, |C-p_x\rangle, |O1-p_x\rangle, |O2-p_x\rangle, |C-p_z\rangle, |O1-p_z\rangle, |O2-p_z\rangle, |C-p_y\rangle, |O1-p_y\rangle,$ and $|O2-p_y\rangle\}$, the *H* is given as

$$
H = \begin{bmatrix}
\varepsilon_s^C & t_1 & t_1 & 0 & t_2 & t_2 & & & & & & \\
t_1 & \varepsilon_s^O & 0 & t_3 & 0 & 0 & & & & & & \\
t_1 & 0 & \varepsilon_s^O & t_3 & 0 & 0 & & \alpha & & & 0 & \\
0 & t_3 & t_3 & \varepsilon_p^C & t_4 & t_4 & & & & & & \\
t_3 & 0 & 0 & t_4 & \varepsilon_p^O & 0 & & & & & & \\
t_3 & 0 & 0 & t_4 & 0 & \varepsilon_p^O & & & & & & \\
 & & & & & & \varepsilon_p^C & t_5 & t_5 & & & \\
 & & \alpha^* & & & & t_5 & \varepsilon_p^O & 0 & & 0 & \\
 & & & & & & t_5 & 0 & \varepsilon_p^O & & & \\
 & & & & & & & & & \varepsilon_p^C & t_5 & t_5 \\
 & & 0 & & & & & 0 & & t_5 & \varepsilon_p^O & 0 \\
 & & & & & & & & & t_5 & 0 & \varepsilon_p^O
\end{bmatrix}. \quad (3)
$$

Here the onsite energies and covalent interactions are represented by $t$ and $\varepsilon$, respectively. The term $\alpha$ represents an off-diagonal $6 \times 3$ block representing the C-$\{s,p_x\}$ – O-$p_z$ interactions.

For the linear molecule, $H$ has an irreducible $6 \times 6$ block representing the axial $\sigma$ interaction among C-$\{s,p_x\}$ and O-$\{s,p_x\}$ orbitals, and an irreducible doubly degenerate $3 \times 3$ block representing the $\pi$ interaction between C-$p_y/p_z$ and O-$p_y/p_z$ orbitals. In this case, elements of $\alpha$ become zero. As shown in Fig. 4(a), corresponding to the $\sigma$ interactions, the resulting six eigenstates in increasing energy order are $3\sigma_g$, $2\sigma_u$, $4\sigma_g$, $3\sigma_u$, $5\sigma_g^*$ and $4\sigma_u^*$ [60,61]. The $\pi$ interactions of the $3 \times 3$ block give rise to three hybridized states, namely, the bonding ($1\pi_u$), antibonding ($2\pi_u^*$) and an occupied ($1\pi_g$) state which is purely of O-$p_{y/z}$ character.

Upon bending in the $xz$ plane, the linear $D_{\infty h}$ symmetry of the $CO_2$ molecule [62] is broken and the O atoms come closer to initiate interactions among their orbitals. Furthermore, the forbidden $p_x$-$p_z$ interactions are now allowed to make $\alpha$ a nonzero block. Thereby, the $12 \times 12$ Hamiltonian matrix is now decomposed into an irreducible $9 \times 9$ block representing the $\sigma$+$\pi$ interactions in the $xz$ plane and an irreducible $3 \times 3$ block representing the $\pi$ interactions along the $y$ axis. The nine eigenstates occupying the $xz$ plane in increasing energy order are $3a_1$, $2b_2$, $4a_1$, $3b_2$, $5a_1$, $4b_2$, $6a_1$, $7a_1$, $5b_2$ and the three eigenstates aligned along $y$ axis are $1b_1$ ($1\pi_u$), $1a_2$ ($1\pi_g$), $2b_1$ ($2\pi_u^*$) [63,64] as shown in Fig. 4(b).

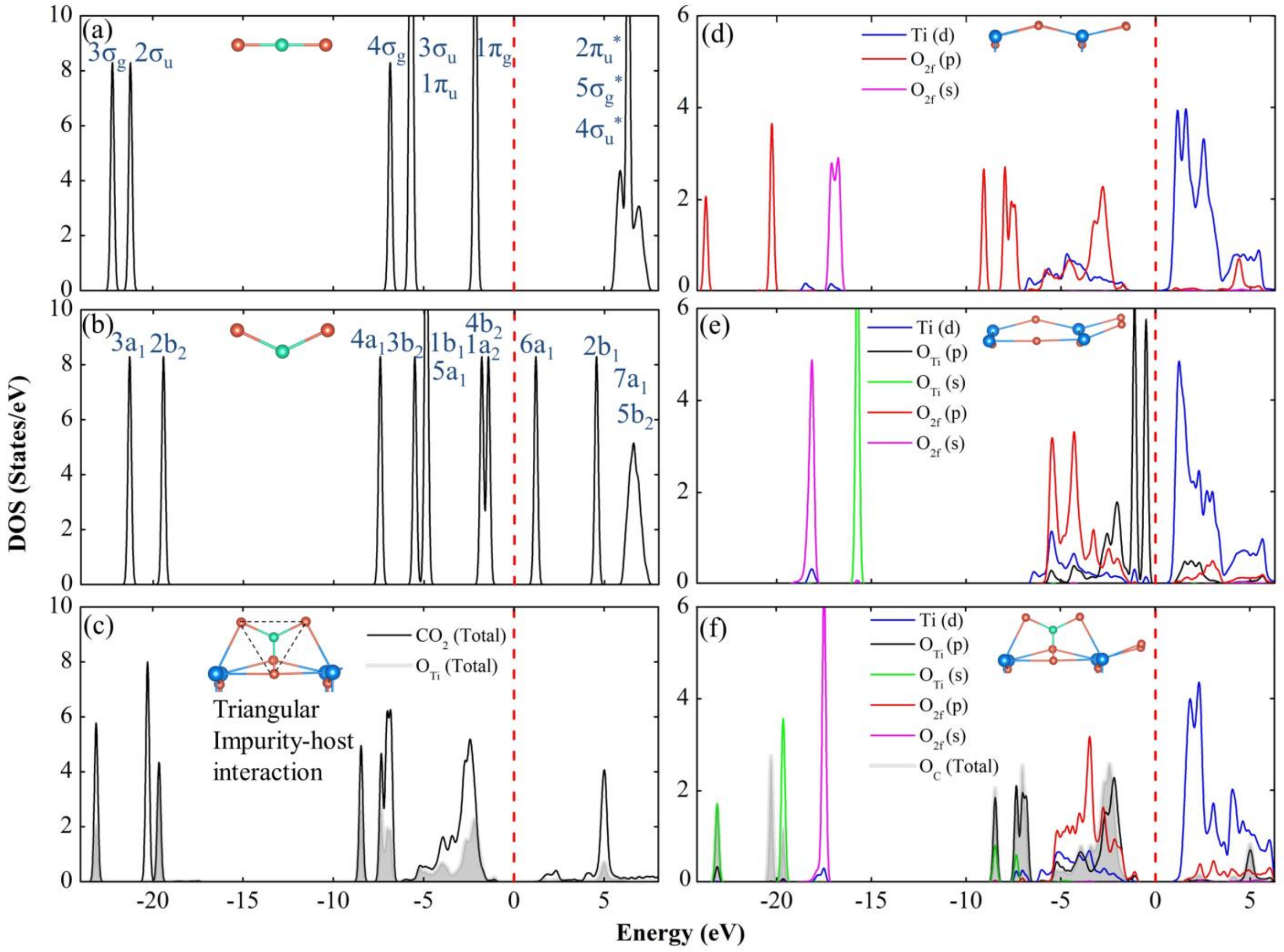

**FIG. 4.** The density of states (DOS) of both $CO_2$ and $TiO_2$ (001) surface before and after adsorption over ontop $O_{2f}$ site. **(**a – c**)** The total density of states of linear, bent and the formed $CO_3^{\delta-}$ complex respectively. (d**)** and (e) The partial DOS of $O_{2f}$-{*s*,*p*} and Ti-*d* states of bare (001) surface without and with distortion. (f) Change in DOS of Ti-*d*, $O_{Ti}$-{*s*,*p*}, $O_C$-*p*, $O_{2f}$-*p* states after adsorption. The molecular states of $CO_2$ are transformed to Bloch states after the interaction. Here, zero separates the occupied and unoccupied level. The symbol * indicates the antibonding states.

For a pristine $TiO_2$ surface, Ti-*d* and $O_{2f}$-*p* states hybridize to form dispersive valence bands dominated by $O_{2f}$-*p* characters and dispersive conduction bands dominated by Ti-*d* characters which are reflected in the partial DOS shown in Fig. 4(d). In the case of a distorted $TiO_2$ surface (state II of Fig. 2), the $O_{Ti}$ atom is pushed downwards by 0.47 Å and the Ti atom moves away from the $O_{Ti}$ atom to bring a significant change in the chemical bonding process. First, the increase in $O_{Ti}$ – Ti bond length makes $O_{Ti}$-*2s* orbitals less participatory in the O-{*s*, *p*} – Ti-*d* covalent interaction and become more localized [Fig. 4(e)]. Therefore, there is a net increase in the occupancy of the $O_{Ti}$-*2s* orbital. Secondly, with the formation of linear Ti – $O_{Ti}$ – Ti geometry (along the *x* axis), except for the weak $O_{Ti}$-$p_z$ – Ti-$d_{xz}$ π interaction, the other $p_z$-*d* interaction vanishes, which can be explained through the Slater-Koster tight-binding matrix elements [65]. As a result, a decrease in the occupancy of the $O_{Ti}$-$p_z$ orbital is observed. Overall, we find a net increment of 0.02 electrons on the $O_{Ti}$ site as discussed in charge-transfer mechanism. Furthermore, due to the breakdown of the translational symmetry, the $O_{Ti}$-$p$ states partially lose their Bloch character and disorder-induced quasilocalized states appear which can be seen from the PDOS of Fig. 4(e).

On examining the situation after the adsorption, we found that the disorder-induced quasilocalized $O_{Ti}$-$p$ states initiate interactions with the molecular quantum states of bent $CO_2$ which can be classified into the nearest-neighbor $O_{Ti}$-{$s$, $p$} – C-{$s$, $p$} and the triangular $O_{Ti}$-{$s$, $p$} – $O_C$-{$s$, $p$} impurity-host interactions. The resulting DOS is shown in Fig. 4(c). Here, we found that $O_{Ti}$-dominated localized states resemble those of the $O_C$ molecular quantum states.

Similarly, the molecular quantum states of $O_C$ are found to interact with $p$-$d$ hybridized Bloch states of the host. Such interactions are supported by the fact that the $O_C$ – Ti bond length (~2.06 Å) is comparable to that of the bulk Ti – $O_{2f}$ bond. Also, through DOS, we observed the emergence of $O_C$-{$s$,$p$} Bloch characters in the valence band spectrum [Fig. 4(f)]. Due to this interaction, the antibonding $6a_1$ state, which has partial $O_C$-$p_z$ character, is now occupied, justifying the charge-transfer mechanism shown in Fig. 3(a). Both structural and DOS characteristics along with the Löwdin charge analysis imply a strong impurity-host (adsorbate-adsorbent) interaction, which leads to the formation of carbonate complex with tridentate coordination.

## B. Anisotropic binding

In previous sections, the total energy calculations and chemical bonding analysis reveal that $O_{2f}$ is the most preferred site for adsorption of $CO_2$ to form a $CO_3^{-0.816}$ complex. Also, we have found that the threefold-coordinated oxygen ($O_{3f}$) does not participate in the impurity-host interaction. Since two consecutive Ti – $O_{2f}$ – Ti chains are separated by a line of $O_{3f}$ atoms [see Fig. 5(a)], the influence of impurity states formed by the $CO_3^{\delta-}$ complex is restricted only to its parent chain. This implies that binding energy cannot be described as a function of coverage alone; rather it is governed by the spatial distribution pattern of $CO_2$ molecules leading to anisotropy.

We have adopted the supercell formalism to vary the coverage and spatial distribution of the adsorbate. Supercells of different size and dimension are considered and allow one $CO_2$ molecule to get adsorbed over each of them. The coverage ($\theta$) is defined as the number of adsorbate molecules per $TiO_2$ formula unit present on the surface. To examine the effect of $\theta$ on the BE for isotropic distribution of the adsorbates, we varied the supercell size from $2 \times 2$ ($\theta = 0.25$) to $5 \times 5$ ($\theta = 0.04$) [Figs. 5(a) and 5(b)]. We observed that the binding strength increases with decrease in coverage [Fig. 5(g)]. The BE at two ends of the coverage spectrum differs significantly by 1.3 eV, as illustrated in Fig. 5(g). Identical θ values can be achieved from anisotropic distribution as well. For example, $\theta$ = 0.25 can be obtained by adsorbing a molecule over $2 \times 2$, $4 \times 1$ and $1 \times 4$ supercell. While the first one represents an isotropic distribution, the other two represent dense distribution along [010] and [100], respectively.

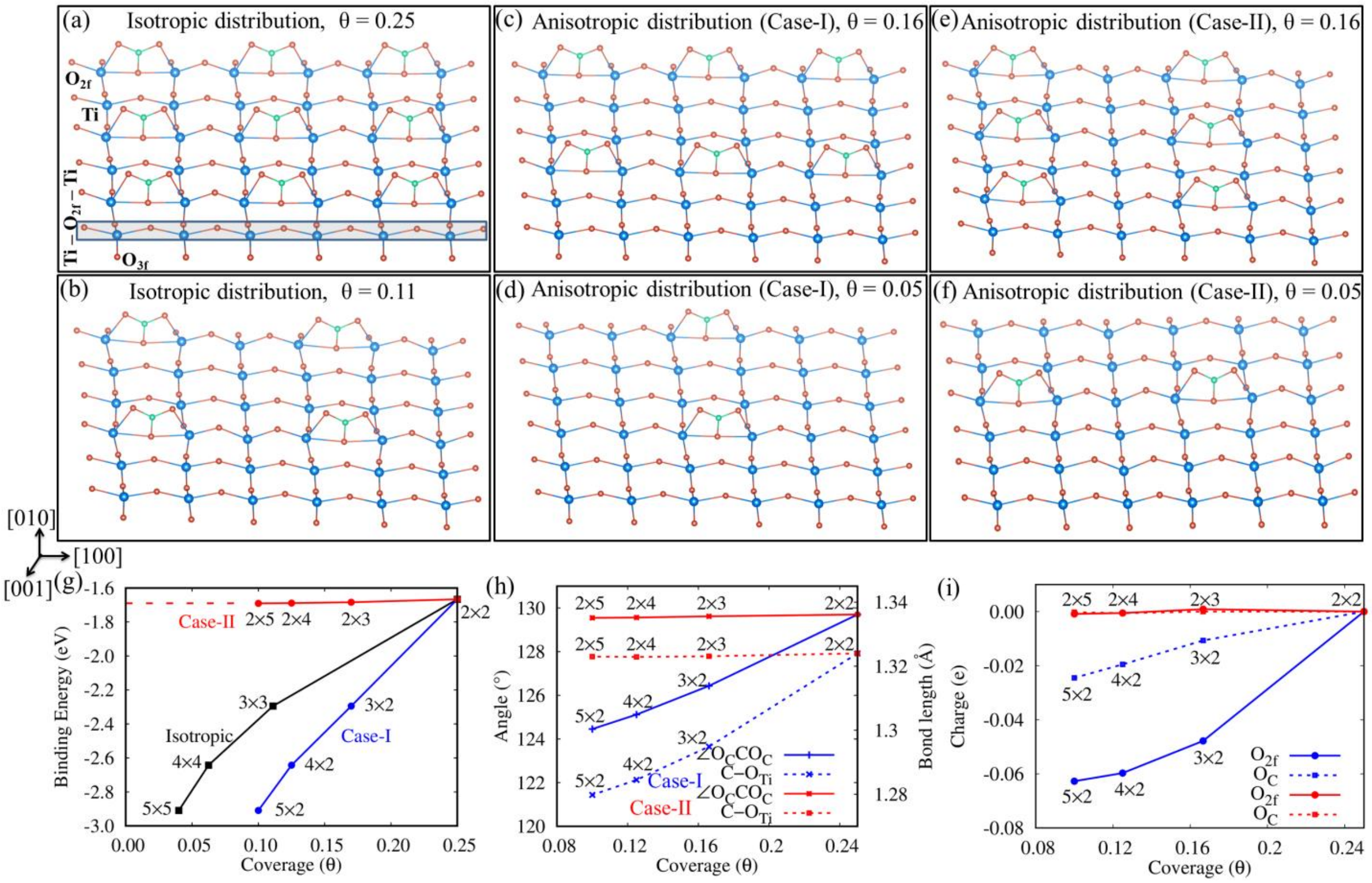

**FIG. 5.** Schematic illustration of isotropic distribution of $CO_2$ molecules for (a) $\theta$ = 0.25 and (b) $\theta$ = 0.11. Similarly, anisotropic distribution is presented through case I (varying distribution along [100]) and case II (varying distribution along [010]). (c) and (d) show distribution as defined in case I with $\theta$ = 0.16 and $\theta$ = 0.05 respectively. (e) and (f) show distribution as defined in case II with $\theta$ = 0.16 and $\theta$ = 0.05 respectively. (g) Variation in BE as a function of coverage with both isotropic (black) and anisotropic (case I: blue; case II: red) distributions. (h) The $\angle O_C C O_C$ bond angle and $C - O_{Ti}$ bond length as a function of $\theta$. (i) Electron accumulation on $O_C$ and $O_{2f}$ atoms due to chemisorption, measured with respect to the isotropic distribution of $\theta$ = 0.25 ($2 \times 2$ supercell), as a function of $\theta$.

In order to analyze the coverage dependent adsorption behavior, for anisotropic distribution, we consider two cases. In case I, we kept the cell size constant along [010] and varied it along [100], i.e., $2 \times 2$, $3 \times 2$, $4 \times 2$ and $5 \times 2$ [Figs. 5(c) and 5(d)]. Similarly, in case II, the cell size remained fixed along [100] and varied along [010], i.e., $2 \times 2$, $2 \times 3$, $2 \times 4$, $2 \times 5$ [Figs. 5(e) and 5(f)]. The calculated BE for both the cases are presented in Fig. 5(g). It can be clearly seen that for case I, binding strength increases from -1.6 eV ($2 \times 2$) to -2.9 eV ($5 \times 2$). It may be noted that a near identical variation in BE is observed [Fig. 5(g)] for isotropic distributions (2 × 2 to 5 × 5). However, for case II, despite an increase in the supercell size (or decrease in coverage), the BE remains constant at -1.6 eV. This confirms that BE depends only on the distance between the adsorbed molecules lying along the parent Ti – $O_{2f}$ – Ti chain which also implies that for a given coverage BE can vary through anisotropic distributions. Further, we found that two neighbor $O_{2f}$ sites along the Ti – $O_{2f}$ – Ti chain cannot chemisorb the $CO_2$ molecules due to a large electrostatic repulsion leading to a positive BE of 6.52 eV. Therefore, the maximum coverage of $\theta = 0.5$ can be achieved on the anatase (001) surface.

To understand the cause of anisotropic behavior in BE, we have examined the structural distortion and Löwdin charges for the aforementioned cases. In case I (defined through n × 2 supercell), on decreasing the coverage, the distortion in the impurity-host complex increases and was found to be more bound. The calculations reveal that the C – $O_{Ti}$ bond length decreases from 1.32 Å to 1.28 Å and the $O_C$ – C – $O_C$ bond angle decreases from 129.70° to 124.45° with decrease in $\theta$ from 0.25 to 0.10 [Fig. 5(h)]. However, for case II (defined through 2 × n supercell), we observed that the extent of distortion is independent of the supercell size (or $\theta$) and hence, the BE remains unchanged. The concurrent effect of distortion in bond length and bond angle of $CO_2$ as well as that of $TiO_2$ surface affects the chemical bonding, which, in turn, leads to the redistribution of charges. Figure 5(i) shows that the charges on the $O_{2f}$ and $O_C$ vary significantly for case I, while they remain the same for case II, which supports the earlier inference that the impurity-host interaction is sensitive only along Ti–$O_{2f}$–Ti chain.

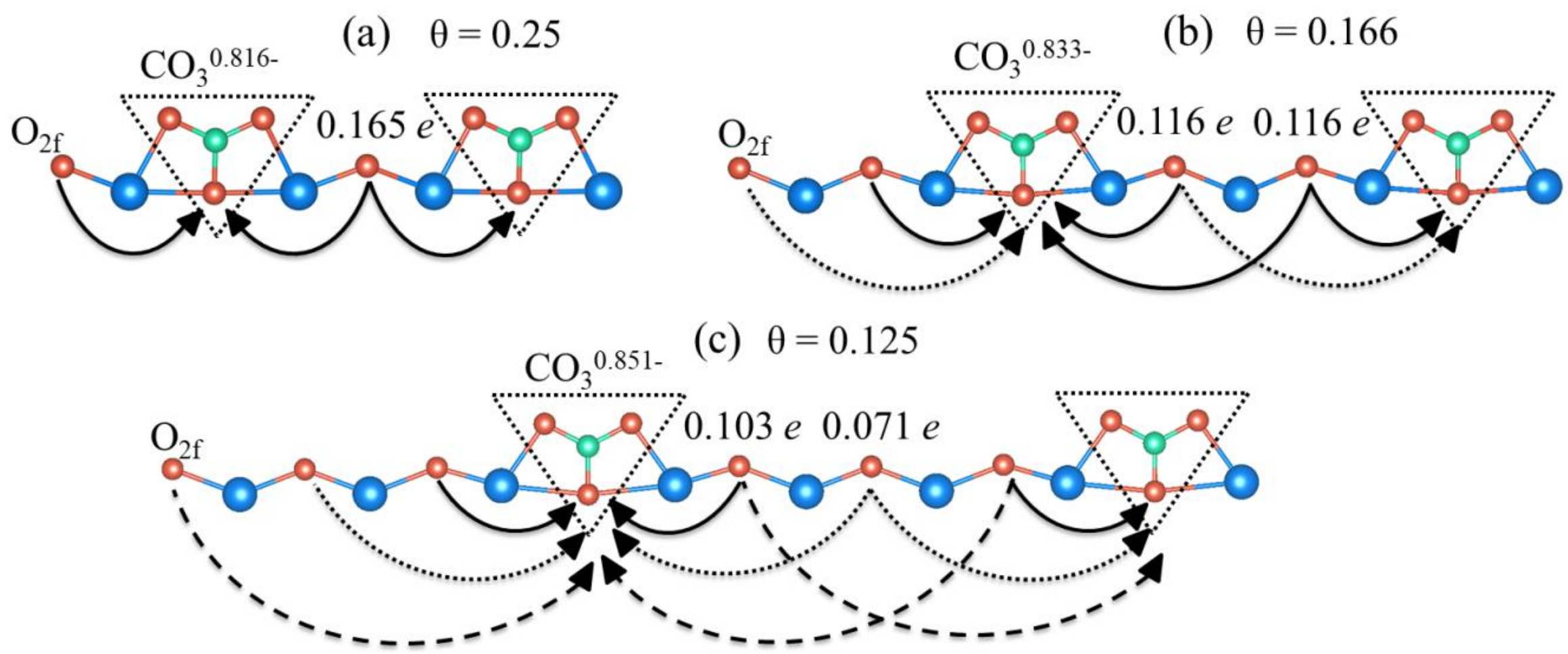

**FIG. 6.** Net charge on the formed carbonate complex and loss of electrons on $O_{2f}$ upon chemisorption. The results presented here are for anisotropic distribution as in case I with (a) $\theta$ = 0.25, (b) $\theta$ = 0.166, and (c) $\theta$ = 0.125. The curved arrows indicate the direction of charge transfer from host to the complex. For low values of θ, far away $O_{2f}$ atoms also contribute to the charge transfer leading to a stronger binding energy.

The reason behind the sharp decrease in BE with decrease in coverage for case I can be explained from the nature of charge transfer between the host and impurity which is illustrated in Fig. 6. When $\theta = 0.25$, the charge transferred ($\Delta$Q) from $O_{2f}$ to the carbonate complex is 0.165 *e* [Fig. 6(a)]. On decreasing $\theta$, the formed carbonate complex ($CO_3^{\delta-}$) become more isolated and large numbers of $O_{2f}$ atoms are available to participate in the impurity-host interaction. Even though the charge transfer from individual $O_{2f}$ is relatively less compared to the case of $\theta = 0.25$, the net charge transfer to the carbonate complex is found to be higher. Therefore, with decrease in coverage, the range of impurity-host interaction as well as the net charge transfer to the carbonate complex increases which lead to a stronger BE. Even though weak, the other interactions that affect the BE are the electrostatic repulsion between the charge clouds of the adjacent carbonate complexes and the increase in elastic energy due to structural distortions. The former strengthens the adsorption for diluted distribution of the $CO_2$ molecules while the later tends to weaken it. The net change in the BE is dominantly affected by the impurity-host interaction. Findings from both the structural and charge analysis complement the observed binding energy behavior [Fig. 5(g)] and also conclude that the direction-dependent chemical bonding is the most dominant factor in causing the BE anisotropy.

### C. Trajectory of randomly oriented $CO_2$ molecules

The fundamental point of the discussion made so far is that the $CO_2$ molecule prefers to be chemisorbed over ontop $O_{2f}$ site with its molecular axis along the [100] axis, i.e., along Ti – $O_{2f}$ – Ti chain. However, in a practical situation, the far away $CO_2$ molecule can have any random orientation ($\Phi$) with respect to [100] axis [Fig. 7(a)]. In such cases, it is necessary to study the trajectory of $CO_2$ leading to the adsorption process. In Fig. 7, we have traced the trajectory for different $\Phi$ values ranging from 0 to $\pi$ with respect to the [100] axis.

As computed earlier, the net charge density on the $TiO_2$ surface is $-2.1\times10^{-3}$ *e*/Å$^2$, whereas the charges on C and $O_C$ atoms of the $CO_2$ molecule are +0.712 *e* and -0.356 *e*, respectively. As a result, the adsorbate feels the electrostatic interaction. Figure 7(b) shows that when the $CO_2$ molecule appears below a critical distance ($d_c$ ~5.5 Å) from the charged $TiO_2$ surface, it experiences an attractive force. When the orientation of the molecular axis is along the preferred alignment [100] ($\Phi = 0$), the net attractive force is along the C – $O_{Ti}$ axis. However, a slight deviation from the preferred alignment ($\Phi \neq 0$) induces a restoring torque about the C – $O_{Ti}$ axis. To understand further, we varied $\Phi$ from 0 to $\pi$ and calculated both the torque and energy. The results are shown in Fig. 7c. The graph, obtained for a separation distance of 2 Å, shows that the torque increases initially and after reaching a critical angle of $\pi/4$, it gradually reduces to zero at $\pi/4$. Moreover, variation in the total energy suggests that perpendicular orientation ($\Phi = \pi/2$) with respect to the preferred alignment leads to an energetically unstable position but with a zero net torque. As a result, even a minimal perturbation triggers a torque and brings it back to the equilibrium position (i.e., along the Ti –$O_{2f}$ –Ti chain). Here, we infer that irrespective of the initial orientation of the $CO_2$ molecule, the electrostatic interaction exerted by the $TiO_2$ surface ensures the alignment of this molecule along the Ti – $O_{2f}$ – Ti chain and thereby, the formation of the carbonate complex.

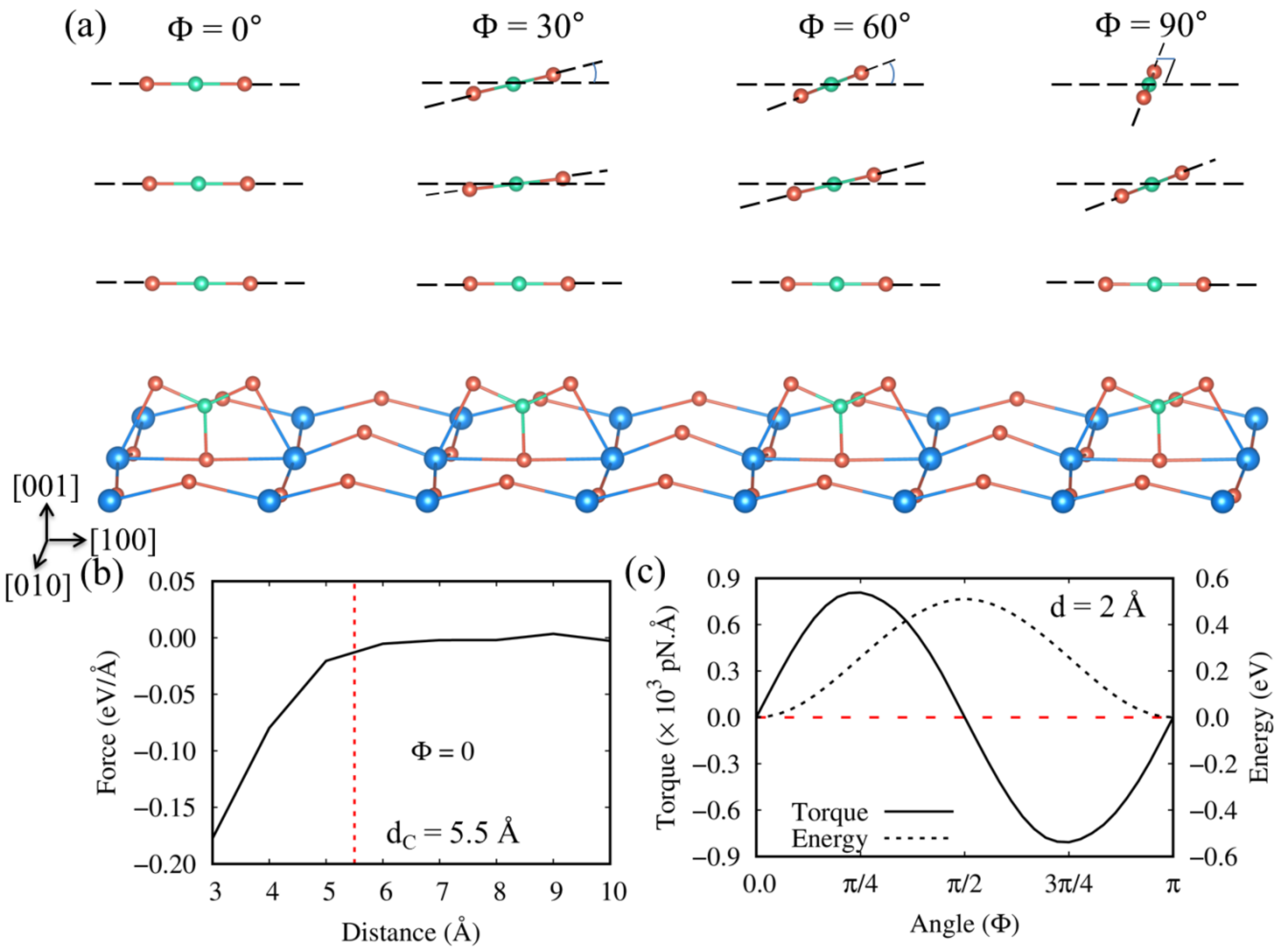

**FIG. 7.** (a) Trajectory of the $CO_2$ molecule, adsorbing over ontop $O_{2f}$, with varied initial orientation, Φ, of molecular axis ranging from 0 to $\pi$. Here, Φ = 0 defines the orientation of molecular axis along [100]. (b) The electrostatic force between the charge cloud of $CO_2$ molecule and $TiO_2$ as a function of the separation distance for Φ= 0. (c) The variation in torque (solid line) and energy (dashed line) as a function of Φ. Here, energies are given with respect to Φ = 0.

## D. Future Scope

The aforementioned discussions collectively conclude that while the $CO_2$ adsorption is sensitive to the spatial distribution of these molecules, it is independent of the initial orientation of the O – C – O axis. These two phenomena offer the scope to tailor the $CO_2$ adsorption, which we have demonstrated through a conceptual experiment as shown in Fig. 8. Here, a virtual nozzle is employed to confine the flow of $CO_2$ molecules along a certain direction ($\psi$) with respect to the Ti – $O_{2f}$ – Ti chain. Assume that large numbers of such nozzles are kept with equispaced separation ($d_{nozzle}$). The concentration of $CO_2$ molecules can be defined in terms of neighboring $CO_2 - CO_2$ separation ($d_{CO_2}$) along the direction of gas flow. On varying these three parameters, $d_{nozzle}$, $d_{CO_2}$, and $\psi$, we can achieve different spatial distributions of $CO_2$ molecules and a few of them are schematically illustrated in Figs. 8(a)-8(e).

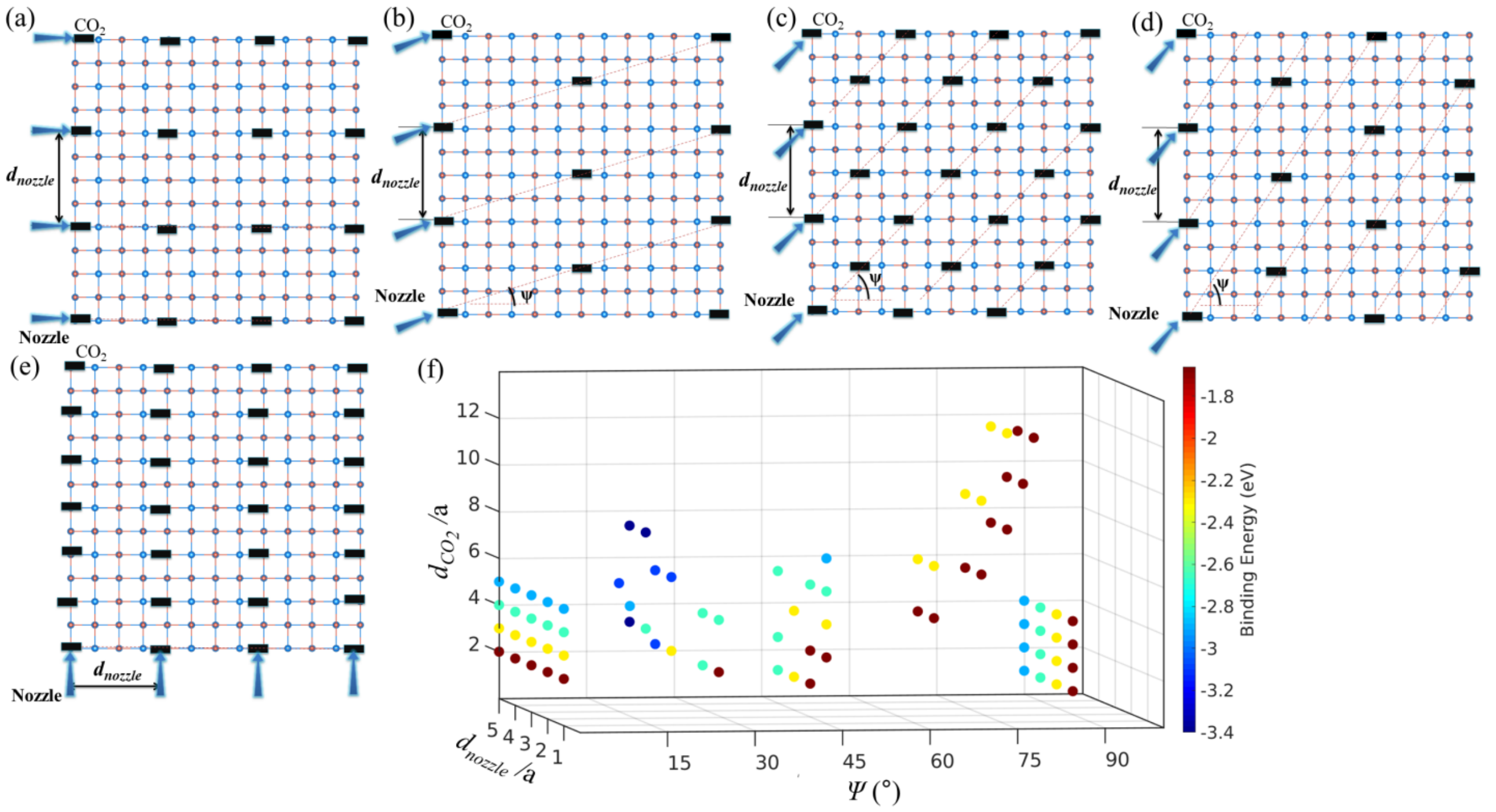

**FIG. 8.** (a – e) Different spatial distribution of $CO_2$ achieved by varying $d_{nozzle}$, $d_{CO_2}$ (distances are in terms of lattice parameter 'a') and $\psi$. (f) Binding energy for varied spatial distribution define the points in a three dimensional space spanned by $d_{nozzle}$, $d_{CO_2}$ and $\psi$. Here, the color gradient signifies the variation in binding energy.

We propose a model that relates the $CO_2$ adsorption behavior to $d_{nozzle}$, $d_{CO_2}$, and $\psi$. As discussed earlier, for chemisorption, the net $CO_2$ coverage on the (001) surface cannot exceed 50 %. Therefore, continuous variation of $d_{nozzle}$, $d_{CO_2}$, and $\psi$ is restricted which leads to certain discrete configurations in these three parameter spaces. As each configuration presents a unique spatial distribution of $CO_2$ molecules, there is a BE variation in these three parameter space which is plotted in Fig. 8(f). As $\psi$ goes to 0 [Fig. 8(a)], the flow is along the Ti – $O_{2f}$ – Ti chain and therefore, rather than the $d_{nozzle}$, the $d_{CO_2}$ determines the BE, which is reflected in Fig. 8(f). For $\psi = \pi/4$, both $d_{nozzle}$ and $d_{CO_2}$ affects the BE. On the other hand, for $\psi = \pi/2$, the flow is along the Ti – $O_{3f}$ – Ti chain and hence, the BE is independent of $d_{CO_2}$. A closer look at any particular configuration suggests that it is the $CO_2 - CO_2$ separation along the Ti – $O_{2f}$ – Ti chain that determines BE. This is in accordance with the BE-anisotropy plots of Fig. 5. Although there are practical challenges, successful experimental implementation of this concept will create new paradigms to engineer the binding energy of $CO_2$ within a range of ~1.5 eV over the anatase $TiO_2$ (001) surface.

## E. $CO_2$ adsorption in the presence of water

To mimic the real experimental conditions, it is desirable to investigate the coadsorption of $CO_2$ and $H_2O$. For this purpose, we have considered a 2×2 supercell of $TiO_2$ surface on which one $H_2O$ and one $CO_2$ molecule are adsorbed. We have adopted the following three ways to study the adsorption process: (i) first $H_2O$ is adsorbed followed by $CO_2$ adsorption, (ii) $CO_2$ is adsorbed followed by $H_2O$ adsorption, and (iii) both the molecules are simultaneously adsorbed. The initial, intermediate and final configurations of each of these cases are shown in Fig. 9. The final configuration is found to be identical irrespective of the steps of the adsorption process, suggesting that we have achieved the global minimum configuration. It is further confirmed by the fact that the net coadsorption energy is found to be same for each of these cases.

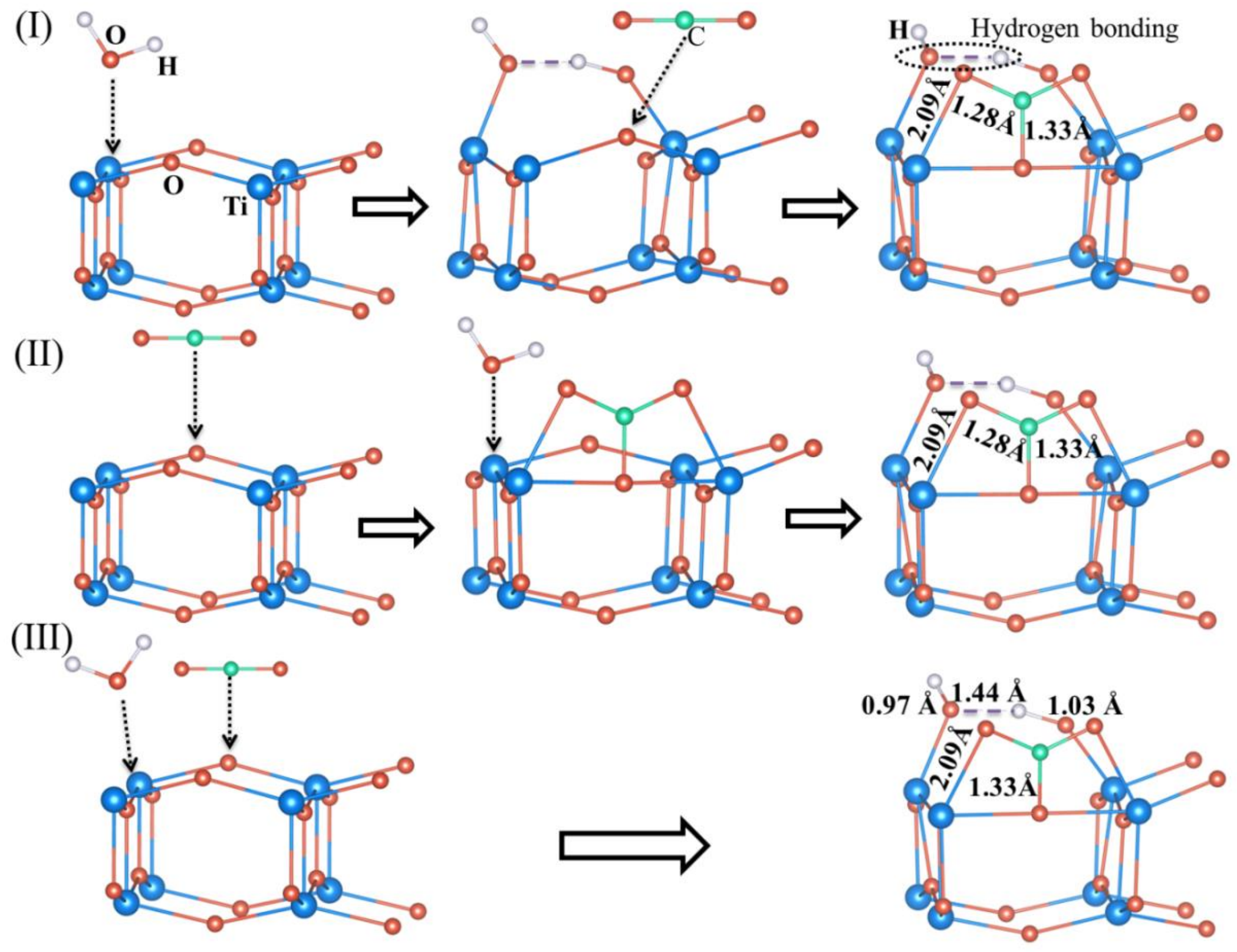

**FIG. 9.** Three possible ways to study the co-adsorption of $CO_2$ and $H_2O$ on anatase $TiO_2$ (001) surface. (I) First $H_2O$ is adsorbed on $TiO_2$ surface followed by $CO_2$ adsorption, (II) $CO_2$ is adsorbed on $TiO_2$ followed by $H_2O$ adsorption and (III) shows the adsorption of $H_2O$ and $CO_2$ simultaneously. The final configurations in three cases are the same. The blue, red, green, and sliver stand for Ti, O, C and H atoms respectively. The H bonding between O and H is indicated by dotted lines.

We analyze the first case in more detail to gain further insight into the adsorption of $CO_2$ in the presence of water. While there are many sites for adsorption, in the minimum energy configuration, the $H_2O$ molecule binds to the surface by forming a bond between surface Ti and $O_W$ (oxygen of water molecule) [see Fig. 10(a)] and a new bond between the neighboring $O_{2f}$ and H is developed. Finally, the $H_2O$ molecule dissociates with the formation of two hydroxyl (OH) groups, which are connected to each other through a hydrogen bond is established between the two OH groups. The adsorption energy of $H_2O$ is calculated to be -2.23 eV. The earlier reported calculations, carried out using the GGA exchange-correlation functional and using only the Γ point of the *k* space, predict the adsorption energy close to -1.7 eV [66,67]. In another pseudopotential and GGA based calculation, a 2×2×1 *k* mesh was used and the resulted adsorption energy was -2.05 eV [68]. As the present calculations use a large (4×4×1) *k* mesh and employ long-range van der Waals corrections, an improvement in the accuracy by ~0.2eV was achieved.

The hydrated $TiO_2$ surface offers multiple sites for $CO_2$ adsorption as can be seen in Fig. 10. Like the case of a pristine surface, the BE graph [Fig. 10(f)] shows that the adsorption at $O_{2f}$ is most favorable. However, with the presence of $H_2O$, the BE strength decreases slightly from 1.66 eV by 0.26 eV. We attribute this difference to the repulsive interaction between the activated carbonate complex and the dissociated hydroxyl ion.

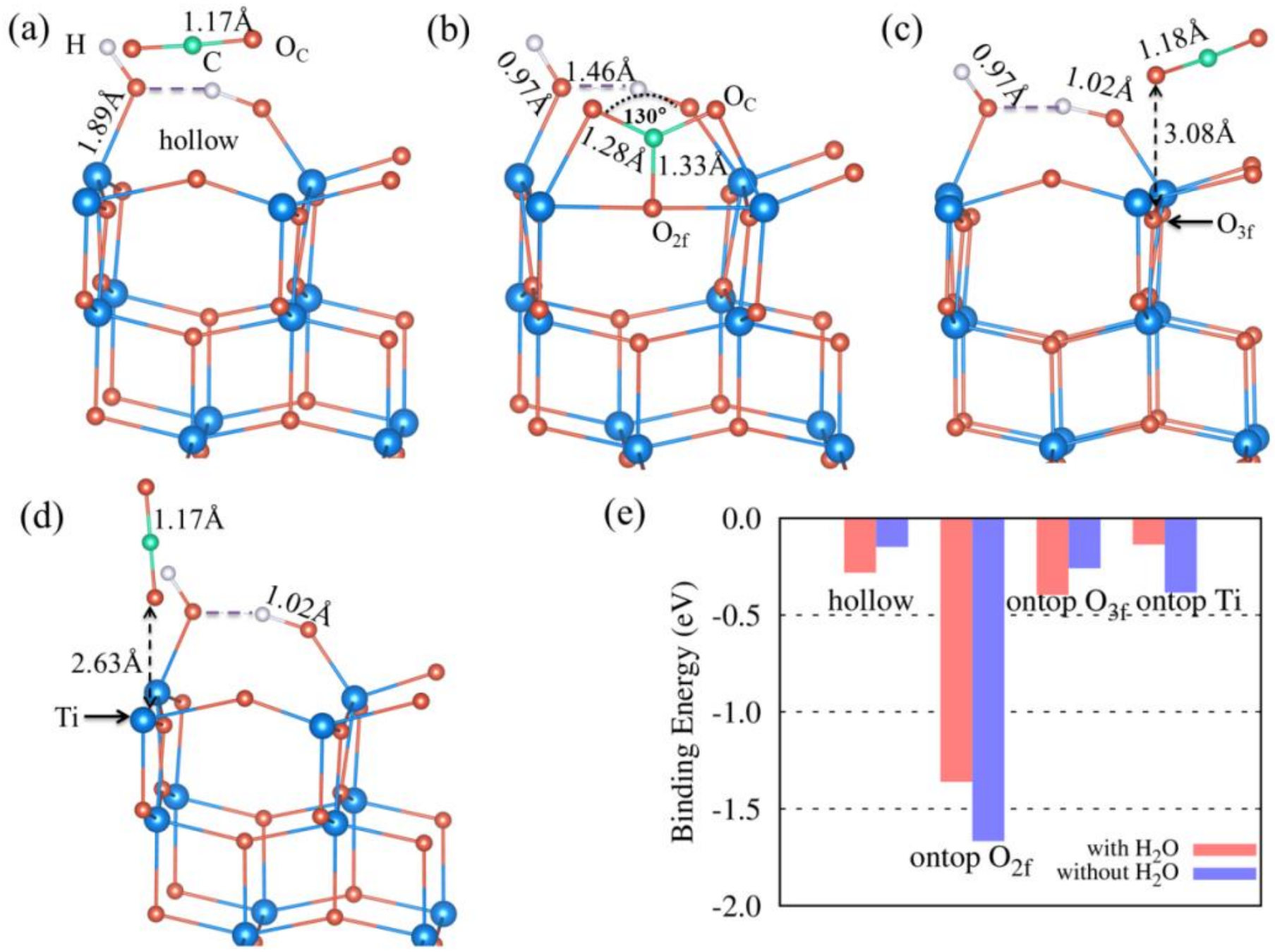

**FIG. 10.** $CO_2$ adsorption on $H_2O$ adsorbed $TiO_2$ surface at (a) hollow site, (b) ontop $O_{2f}$, (c) ontop $O_{3f}$ and (d) ontop Ti. (f) The BE barplot for $CO_2$ adsorption with and without $H_2O$ adsorbed $TiO_2$.

The coadsorbed stable structure, i.e. with two hydroxyls and one active $CO_3$ complex, can make a transition to form hydrocarbon complexes. The first step in this direction is the formation of the bicarbonate $HCO_3$ complex [69,70]. The migration of H from the hydroxyl group to the carbonate complex leads to the formation of surface bicarbonate (-$HCO_3$). From the CI-NEB calculations, the activation barrier is calculated to be 0.89 eV. The gradual transition is demonstrated in Fig. 11.

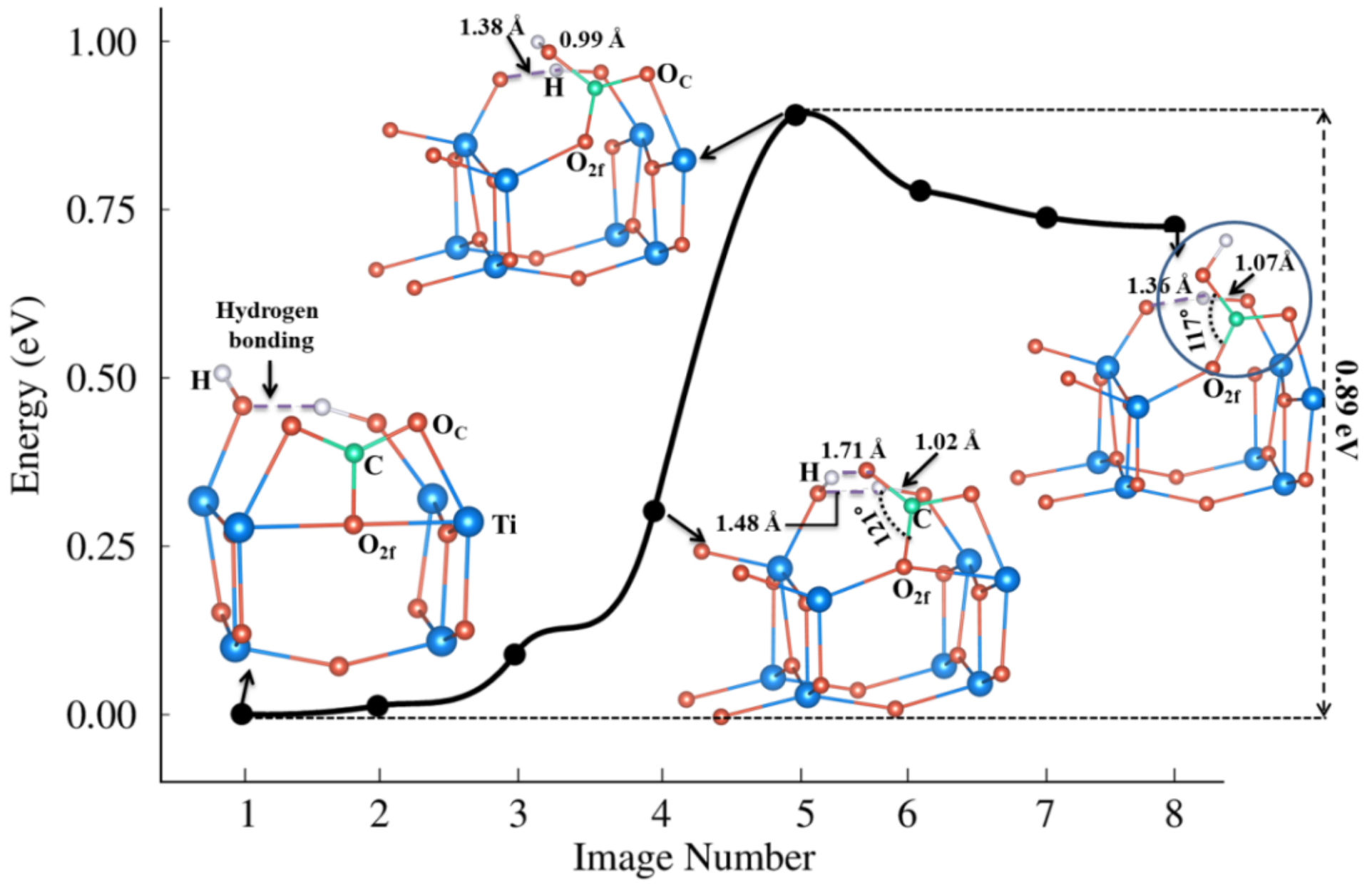

**FIG. 11.** Reaction pathways for transition of a $CO_3$ complex to a $HCO_3$ complex in presence of $H_2O$ molecule as obtained from CI-NEB simulations.

## IV. CONCLUSIONS

In summary, from comprehensive density functional calculations and molecular-orbital theory, we developed a three-state model to explain the mechanism of $CO_2$ adsorption on the most reactive anatase $TiO_2$ (001) surface. We found that the charge transfer between the host and adsorbate occurs only along the chain consisting of alternately placed Ti and twofold oxygen atoms (Ti – $O_{2f}$ –Ti) on the surface. Therefore, there is a binding energy anisotropy which suggests that spatial distribution of $CO_2$, rather than the coverage as reported earlier, is the deciding factor in determining the adsorption behavior. In fact, our binding energy analysis reveals that a maximum of 50% coverage can be achieved for chemisorption, beyond which the $CO_2$ molecules experience a repulsive force at the $O_{2f}$ site. Exploiting the binding energy anisotropy, we propose a conceptual experiment which suggests that a binding energy can be varied in an energy window of about ~1.5 eV by controlling the direction and concentration of gas flow. It may be noted that such binding energy anisotropy are not present in the most stable anatase (101) surface. In the case of co-adsorption of $H_2O$ and $CO_2$, which generally happens in experiments, we find that $H_2O$ dissociates to form two hydroxyl ions. Further, by overcoming a potential barrier of ~0.9 eV, the carbonate complex can break a hydroxyl ion to create the hydrocarbon ($HCO_3$) through the freed H atom.

## ACKNOWLEDGEMENTS

The authors would like to thank HPCE, IIT Madras for providing the computational facility. This work is supported by Defence Research and Development Organization, India through Grant No. ERIP/ER/RIC/201701009/M/01.